\documentclass[aps,prd,nofootinbib,twocolumn,floatfix,superscriptaddress]{revtex4-1}
\usepackage{setspace}
\usepackage{graphicx}
\usepackage{dcolumn}
\usepackage{multirow}
\usepackage{subfigure}
\usepackage{times,mathptm}
\usepackage{float}
\usepackage{color}
\usepackage{amsmath,amsfonts}
\usepackage{mathptmx}
\usepackage{mathrsfs}
\usepackage{bbm}
\usepackage{bm}
\usepackage{xfrac}

\newcommand{\beq}{\begin{equation}}
\newcommand{\eeq}{\end{equation}} 
\newcommand{\bea}{\begin{eqnarray}}
\newcommand{\eea}{\end{eqnarray}}

\renewcommand{\d}{\delta}

\renewcommand{\l}{\lambda}

\newcommand{\Dt}{{\cal D}}

\newcommand{\tK}{\widetilde{K}}

\renewcommand{\b}{\beta}
\renewcommand{\a}{\alpha}

\newcommand{\tr}{\text{Tr}}

\newcommand{\bx}{\mathbf{x}}

\newcommand{\vx}{{\vec{x}}}
\newcommand{\vy}{{\vec{y}}}

\newcommand{\vk}{{\vec{k}}}

\newcommand{\m}{\mu}

\newcommand{\D}{\Delta}

\renewcommand{\th}{\theta}

\newcommand{\oh}{\frac{1}{2}}

\newcommand{\on}{\frac{1}{9}}
\newcommand{\dg}{\dagger}
\newcommand{\non}{\nonumber}

\newcommand{\rf}[1]{(\ref{#1})}
\newcommand{\ra}{\rightarrow}
\newcommand{\pa}{\partial}
\renewcommand{\vec}[1]{\bm #1}

\usepackage{ulem}

\begin{document}

\title{Relative weights approach to SU(3) gauge theories with dynamical fermions at finite density} 
 
\author{Jeff Greensite}
\affiliation{Physics and Astronomy Department, San Francisco State
University,   \\ San Francisco, CA~94132, USA}
\bigskip
\author{Roman H\"ollwieser}
\affiliation{Department of Physics, New Mexico State University, \\
Las Cruces, NM, 88003-0001, USA}
\affiliation{Institute of Atomic and Subatomic Physics, Vienna University of Technology, \\ 
Operngasse 9, 1040 Vienna, Austria}
\date{\today}
\vspace{60pt}
\begin{abstract}

\singlespacing

   We derive effective Polyakov line actions for SU(3) gauge theories with staggered dynamical fermions, for a small sample of lattice couplings, lattice actions, and lattice extensions in the time direction.  The derivation is via the method of relative weights, and the theories are solved at finite chemical potential by mean field theory.  We find in some instances that the long-range couplings in the effective action are very important to the phase structure, and that these couplings are responsible for long-lived metastable states in the effective theory.  Only one of these states corresponds to the underlying lattice gauge theory.
 
\end{abstract}

\maketitle

\singlespacing
\section{\label{intro}Introduction}

   One approach to understanding the phase structure of QCD at finite densities is to map the theory onto a simpler theory, described by an effective Polyakov line action, and then to solve for the phase structure of that theory by whatever means may be available.  At strong couplings and heavy quark masses the effective theory can be obtained by a strong-coupling/hopping parameter expansion, and such expansions have been carried out to rather high orders \cite{Langelage:2014vpa}.  These methods do not seem appropriate for weaker couplings and light quark masses, and a numerical approach of some kind seems unavoidable.  There are, of course, methods aimed directly at the lattice gauge theory, bypassing the effective theory.  These include the Langevin equation \cite{Sexty:2013ica} and Lefshetz thimbles \cite{Scorzato:2015qts}.  In this article, however, we are concerned with deriving the effective Polyakov line action numerically, and solving the resulting theory at non-zero chemical potential by a mean field technique.  In the past we have advocated a ``relative weights'' method  \cite{Greensite:2014isa,Greensite:2013yd}, reviewed below, to obtain the effective theory, but thus far this method has only been applied to pure gauge theory, and to gauge theory with scalar matter fields.  Here we would like to report some first results for SU(3) lattice gauge theory coupled to dynamical staggered fermions.\footnote{For an interesting alternative approach to determining the PLA by numerical means, so far applied to pure SU(3) gauge theory, see 
\cite{Bergner:2015rza}.}
   
\section{The Relative Weights Method}
   
     The effective Polyakov line action (henceforth ``PLA'') is the theory obtained by integrating out all degrees of freedom
of the lattice gauge theory, under the constraint that the Polyakov line holonomies are held fixed.  It is convenient to implement this constraint in temporal gauge (${U_0(\bx,t\ne 0)=\mathbbm{1}}$), so that
\bea
\lefteqn{\exp\Bigl[S_P[U_{\vx},U^\dg_{\vx}]\Bigl] }& & \non \\
&=&  \int  DU_0(\vx,0) DU_k  D\phi \left\{\prod_{\vx} \d[U_{\vx}-U_0(\vx,0)]  \right\} e^{S_L} \ ,
\label{S_P}
\eea
where $\phi$ denotes any matter fields, scalar or fermionic, coupled to the gauge field, and $S_L$ is the SU(3) lattice action (note that we adopt a sign convention for the Euclidean action such that the Boltzman weight is proportional to $\exp[+S]$).  To all orders in a strong-coupling/hopping parameter expansion, the relationship between the PLA at zero chemical potential 
$\m=0$, and the PLA corresponding to a lattice gauge theory at finite chemical potential, is given by
\bea
     S_P^\m[U_\vx,U^\dg_\vx] =  S_P^{\m=0}[e^{N_t \m} U_\vx,e^{-N_t \m}U^\dg_\vx]   \ .
\label{convert}
\eea
So the immediate problem is to determine the PLA at $\m=0$.

      The relative weights method can furnish the following information about $S_P$:  Let $\cal{U}$ denote the space of all Polyakov line (i.e.\ SU(3) spin) configurations $U_\vx$ on the lattice volume.  Consider any path through this configuration space $U_\vx(\l)$ parametrized by $\l$.  Relative weights  enables us to compute the derivative of the effective action $S_P$ along the path
\beq
            \left( {d S_P \over d \l}  \right)_{\l=\l_0}
\eeq
at any point $\{U_\vx(\l_0)\} \in \cal{U}$.  The strength of the method is that it can determine such derivatives along any path, at any point in configuration space, for any set of lattice couplings and quark masses where Monte Carlo simulations can be applied.  The drawback is that it is not straightforward to go from derivatives of the action to the action itself, and in general one must assume some (in general non-local) form for the effective action, and use the relative weight results to determine the parameters that appear in that action. 
    
   In practice the procedure is as follows.  Let
\bea
U'_\vx = U_\vx(\l_0 + \oh \D \l)  ~~~,~~~
U''_\vx = U_\vx(\l_0 - \oh \D \l ) ~~~,
\eea
denote two Polyakov line configurations that are nearby in $\cal{U}$,
with $S'_L,S''_L$ the lattice actions with timelike links $U_0(\vx,0)$ on a $t=0$ timeslice held fixed to  $U_0(\vx,0)=U'_\vx$
and $U_0(\vx,0)=U''_\vx$ respectively.  Defining
\bea
           \D S_P = S_P[U'_\vx] - S_P[U''_\vx] \ ,
\eea      
we have from \rf{S_P},
\bea
e^{\D S_P} &=&  {\int  DU_k  D\phi ~  e^{S'_L} \over \int  DU_k  D\phi ~  e^{S''_L} }
\non \\ 
&=& {\int  DU_k  D\phi ~  \exp[S'_L-S''_L] e^{S''_L} \over \int  DU_k  D\phi ~  e^{S''_L} }
\non \\
&=& \Bigl\langle  \exp[S'_L-S''_L] \Bigr\rangle'' \ ,
\eea
where $\langle ... \rangle''$ indicates that the expectation value is to be taken in the probability measure
\bea
{e^{S''_L} \over  \int  DU_k  D\phi ~  e^{S''_L} } \ .
\eea
This expectation value is straightforward to compute numerically, and from the logarithm we determine $\D S_P$.
Then 
\beq
           \left( {d S_P \over d \l}\right)_{\l=\l_0} \approx {\D S_P \over \D \l} 
\eeq  
is the required derivative.

    The PLA inherits, from the underlying lattice gauge theory, a local symmetry under the transformation 
$U_\vx \ra g_\vx U_\vx g^{-1}_\vx$, which implies that the PLA can depend only on the eigenvalues of the
Polyakov line holonomies $U_\vx$.  Let us define the term ``Polyakov line'' in an SU($N$) theory to refer to the trace of the Polyakov line holonomy
\beq
             P_\vx \equiv  {1 \over N} \tr[U_\vx] \ .
\label{P}
\eeq
The SU(2) and SU(3) groups are special in the sense that $P_\vx$ contains enough information to determine the eigenvalues of $U_\vx$ providing, in the SU(3) case, that $P_\vx$ lies in a certain region of the complex plane.  Explicitly, if we denote the eigenvalues of $U_\vx$ as $\{e^{i\th_1},e^{i\th_2},e^{-i(\th_1+\th_2)}\}$, then $\th_1,\th_2$ are determined by separating \rf{P} into its real and imaginary parts, and solving the resulting transcendental equations 
\bea
  \cos(\th_1) + \cos(\th_2) + \cos(\th_1+\th_2) &=& 3 \text{Re}[P_x] \ ,
\non \\
  \sin(\th_1) + \sin(\th_2) - \sin(\th_1+\th_2) &=& 3 \text{Im}[P_x] \ .
\label{eigenvalues}
\eea
In this sense the PLA for SU(2) and SU(3) lattice gauge theories is a function of only the 
Polyakov lines $P_\vx$.

    We therefore compute the derivatives of the effective action, by the relative weights method, with respect to the
Fourier (``momentum'') components $a_\vk$ of the Polyakov line configurations
\beq
             P_\vx = \sum_\vk  a_\vk e^{i \vk \cdot \vx} \ ,
\eeq
The procedure is to run a standard Monte Carlo simulation, generate a configuration of Polyakov line holonomies  $U_\vx$, and compute the Polyakov lines $P_\vx$.  We then set the momentum mode $a_\vk=0$ in this configuration to zero, resulting in the modified configuration $\widetilde{P}_\vx$, where
\beq
             \widetilde{P}_\vx =  P_\vx - \left({1\over L^3} \sum_\vy P_\vy e^{-i\vk \cdot \vy}\right) e^{i\vk \cdot \vx} \ .
\eeq
Then define
\bea
            P''_\vx &=& \Bigl(\a - \oh \D \a \Bigr) e^{i\vk \cdot \vx} + f \widetilde{P}_x \ ,
\non \\
            P'_\vx &=& \Bigl(\a + \oh \D \a \Bigr) e^{i\vk \cdot \vx} + f \widetilde{P}_x \ ,
\eea
where $f$ is a constant close to one.  We derive the eigenvalues of the corresponding holonomies $U''_x$ and $U'_x$,
whose traces are $P''_\vx,P'_\vx$ respectively, by solving \rf{eigenvalues}.  The holonomies themselves can be taken to be diagonal matrices, without any loss of generality, thanks to the invariance under $U_\vx \ra g_\vx U_\vx g^{-1}_\vx$ noted above.  If we could take $f=1$, then in
creating $P''_\vx,P'_\vx$ we are only modifying a single momentum mode of the Polyakov lines of a thermalized
configuration.  However, there are two problems with setting $f=1$.  The first is that at $f=1$ and finite $\a$ there are usually some lattice sites where $|P'_\vx|,|P''_\vx| > 1$, which is not allowed.
In SU(3) there is the further problem that at some sites the transcendental equations \rf{eigenvalues} have no solution
for real angles $\th_1,\th_2$.  So we are forced to choose $f$ somewhat less than one; in practice we have used
$f=0.8$.  The choice $f=1$ is only possible in the large volume, $\a \ra 0$ limit.   

    From the holonomy configurations $U''_x, U'_x$ we compute derivatives of $S_P$, as described above, with respect to the real part $a^R_\vk$ of the Fourier components $a_\vk$.
    
\section{A heavy-quark ansatz for $\mathbf S_P$}

   The problem is to derive $S_P$ from the derivatives $\pa S_P/\pa a^R_\vk$.  Unfortunately there is no systematic procedure for doing this, and an ansatz for the effective action is required.  For pure gauge theories we have assumed
a bilinear effective action of the form
\bea
             S_P &=& \sum_{\vx \vy} P_\vx P^\dg_\vy K(\vx-\vy) 
\non \\            
                  &=& \sum_\vk a_k a^*_k \tK(\vk) \ ,
\label{pureS}
\eea
where 
\beq
             K(\vx-\vy) = {1\over L^3} \sum_\vk \tK(k) e^{-\vk \cdot (\vx-\vy)} \ .
\eeq
This non-local coupling can be obtained from derivatives 
\beq
            {1\over L^3}\left( {\pa S_P \over \pa a^R_{\vk}}\right)_{a_\vk = \a} = 2 \tK(\vk) \a \ .
\label{observable}
\eeq   
computed by the relative weights method.  A test of the method and the ansatz \rf{pureS} is to compare the
Polyakov line correlator
\beq
     G(R) = \langle P(\vx) P^\dg(\vy) \rangle ~~~,~~~ R=|\vx-\vy|
\eeq
computed in the effective theory with the corresponding correlator computed in the underlying lattice gauge theory.
Excellent agreement was found in SU(2) and SU(3) pure gauge and gauge-Higgs theories \cite{Greensite:2014isa, 
Greensite:2013bya,Greensite:2013yd}.

    Now we are interested in adding dynamical fermions, which break global center symmetry explicitly in the underlying lattice gauge theory, and the problem is to determine their contribution to the effective action.  For heavy quarks the
answer is known \cite{Fromm:2011qi}, and if we denote by $S_F$ the center symmetry-breaking piece of the effective action, then to leading order in the hopping parameter expansion, at non-zero chemical potential, we have
\bea
\exp[S_F(\mu)] = \sum_\vx \det[1+he^{\mu/T}\tr U_\vx]^p\det[1+he^{-\mu/T}\tr U^\dg_\vx]^p
\non \\
\label{eq:hop}
\eea
where determinants can be expressed entirely in terms of Polyakov line operators, using the
identities
\begin{widetext}
\bea
\det[1+he^{\mu/T}\tr U_\vx]&=&1+he^{\mu/T}\tr U_\vx+h^2e^{2\mu/T}\tr
	U^\dg_\vx+h^3e^{3\mu/T},\non\\
\det[1+he^{-\mu/T}\tr U^\dg_\vx]&=&1+he^{-\mu/T}\tr U^\dg_\vx+h^2e^{-2\mu/T}\tr U_\vx+h^3e^{-3\mu/T} \ ,
\label{eq:hop1}
\eea
\end{widetext}
and where $h = (2\kappa)^{N_t}$, with $\kappa$ the hopping parameter for Wilson fermions,
or $\kappa=1/2m$ for staggered fermions, and $N_t$ is the extension of the lattice in the
time direction.  The power is $p = 1$ for four flavors of staggered fermions, and $p = 2N_f$ for
$N_f$ flavors of Wilson fermions.  It is possible to compute higher order terms in $h$ in a combined strong-coupling/hopping parameter expansion \cite{Langelage:2014vpa}, and of course fermion loops which do not wind around the periodic time direction will also contribute to the center symmetric part of the effective action.

   Our proposal is to fit the relative weights data to an ansatz for $S_P$ based on the massive quark effective 
action, i.e.
\begin{widetext}
\bea
S_P[U_\vx] &=& \sum_{\vx,\vy} P_\vx K(x-y) P_\vy
+p \sum_\vx \bigg\{ \log(1+he^{\mu/T}\tr[U_\vx]+h^2e^{2\mu/T}\tr[U_\vx^\dagger]+h^3e^{3\mu/T}) \non \\
& & \qquad +\log(1+he^{-\mu/T}\tr[U_\vx]+h^2e^{-2\mu/T}\tr[U_\vx^\dagger]+h^3e^{-3\mu/T}) \bigg\}
\label{eq:SP}
\eea
\end{widetext}
where both the kernel $K(\vx-\vy)$ and the parameter $h$ are to be determined by the relative weights method.  The full action is surely more complicated than this ansatz; the assumption is that these terms in the action are dominant, and the effect of a lighter quark mass is mainly absorbed into the parameter $h$ and kernel $K(\vx-\vy)$.
We are aware that this is a strong assumption.  There are two modest checks, however.  First we can compare, at $\m=0$, the Polyakov line correlators computed in the effective theory and the underlying gauge theory, and see how well they agree.  Secondly, if it turns out that the $h$-parameter is very small even for quark masses which are fairly light in lattice units, then that is an indication that more complicated center symmetry-breaking terms are smaller still, and likely to be unimportant, at least at $\m=0$. Finally, we do know qualitatively that an ansatz of this form satisfies the Pauli principle, in that the number density $n$ of quarks per site will saturate, as $\m \ra \infty$, at the correct integer, which is $n=3$ for three colors of staggered unrooted ($p=1$) fermions.  For these reasons we regard the ansatz \rf{eq:SP} as a reasonable starting point for the relative weights approach, to be modified as necessary.

   Components of the wavevector $k_i = 2\pi m_i/L$ are specified by a triplet of integer mode numbers 
$(m_1,m_2,m_3)$, and in this work  we have used triplets
\bea 
&& (000),(100),(110),(200),(210),(300),(311),(400),
\non\\
&& (322),(430),(333),(433),(443),(444),(554)
\label{eq:ks}
\eea
with lattice extension $L=16$ in the three space directions.  In calculating the center symmetry-breaking parameter $h$ and momentum-space kernel $\tK(\vk)$ at $\vk=0$, we gain precision by carrying out the relative weights calculation at imaginary chemical potential $\m/T = i\th$.  This is achieved by constructing $U'_\vx, U''_\vx$ as described above, and then
making the replacements
\bea
U'(\vx,0) &=& e^{i\th} U'_\vx ~,~ U'^\dg(\vx,0) = e^{-i\th} U'^\dg_\vx
\non \\
U''(\vx,0) &=& e^{i\th} U''_\vx ~,~ U''^\dg(\vx,0) = e^{-i\th} U''^\dg_\vx
\eea
which are held fixed in the Monte Carlo simulation.  The simulations are carried out for unrooted staggered fermions, corresponding to $p=1$ in the heavy quark ansatz  \rf{eq:SP}.
The derivative of $S_P$ in \rf{eq:SP} with respect to the real part $a^R_0$ of the Polyakov line zero mode is then
\begin{widetext}
\bea
{1\over L^3} \left( {\pa S_P \over \pa a^R_0} \right)_{a_0=\a}^{\m/T=i\th} &=& 2 \tK(0) \a 
 + \bigg\{(3 h e^{i\th} + 3h^2 e^{2i\th} ) {1\over L^3}\sum_\vx Q_\vx^{-1}(\th) + \mbox{c.c} \bigg\}
\eea
\end{widetext}
where
\beq
          Q_x(\th) = 1 + 3 he^{i\th} P_\vx+  3h^2 e^{2i\th} P_\vx^\dg + h^3 e^{3i\th}
\label{Q}
\eeq
If $h \ll 1$, so that it is consistent to drop terms of $O(h^2)$ and higher, then the derivative simplifies to
\bea
{1\over L^3} \left( {\pa S_P \over \pa a^R_0} \right)_{a_0=\a}^{\m/T=i\th} &=& 2 \tK(0) \a + 6 h \cos \th
\label{deriv1}
\eea
The left hand side of this equation is computed numerically, for a variety of $\a, \th$, by the relative weights
technique.  Plotting the data vs.\ $\a$ at $\th=0$, we can find $\tK(0)$ and $h$ from the slope and intercept, respectively.
However, a more accurate estimate of $h$ is obtained by plotting the results vs.\ $\th$, at fixed $\a$, and then
extrapolating to $\a \ra 0$.  

    Having computed $h$ and $\tK(0)$, the next thing to do is to compute the kernel $\tK(\vk)$ at $\vk \ne 0$, and for this we can set the chemical potential to zero.  We then have the derivative of the action with respect to non-zero modes 
$a_\vk$ of the Polyakov lines
\bea
 {1\over L^3}&& \left( {\pa S_P \over \pa a^R_\vk} \right)_{a_\vk=\a} = 2 \tK(\vk) \a
\non \\ 
&&+    {p\over L^3} \sum_\vx \left( {3h e^{i \vk \cdot \vx} + 3h^2  e^{-i \vk \cdot \vx}\over Q_\vx(0)} + \mbox{c.c} \right)
\eea
Again dropping terms of order $h^2$ and higher, this simplifies to
\bea
 {1\over L^3} \left( {\pa S_P \over \pa a^R_\vk} \right)_{a_\vk=\a} = 2 \tK(\vk) \a
\label{deriv2}
\eea
The left-hand side is computed via relative weights at a variety of $\a$, and plotting those results vs.\ $\a$, $K(\vk)$
is determined from the slope.

\begin{figure}[t!]
\centerline{\scalebox{0.6}{\includegraphics{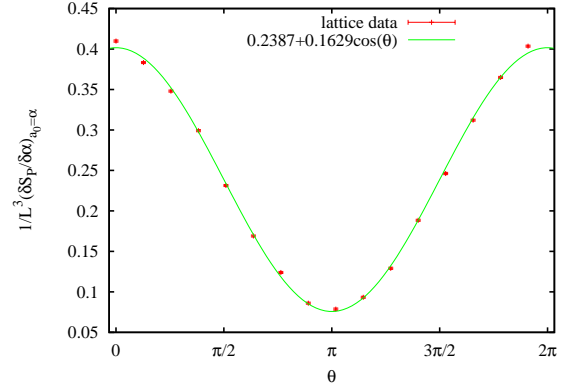}}}
\caption{Derivative of the PLA $\pa S_P/\pa a^R_0$ with respect to the zero momentum component
of the Polyakov lines, evaluated at $a_0=\a=0.03$, vs.\ imaginary chemical potential $\th=\m/T$.  This is for an
underlying lattice gauge theory with a Wilson action at $\b=5.2,~ma=0.35,~N_t=4$. }
\label{dStheta3k0}
\end{figure} 

\begin{figure}[t!]
\centerline{\scalebox{0.6}{\includegraphics{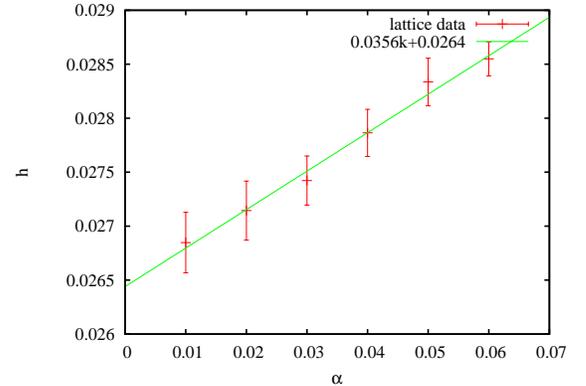}}}
\caption{Parameter $h$ extracted from relative weights data with lattice parameters as in the previous figure, for a variety of $a_0=\a$ values, and extrapolated to $\a=0$.}
\label{h}
\end{figure} 

\begin{figure}[t!]
\centerline{\scalebox{0.6}{\includegraphics{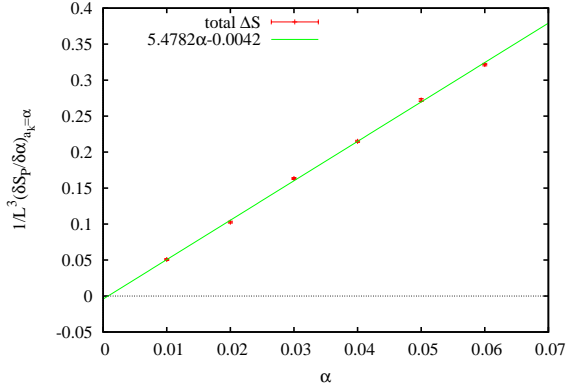}}}
\caption{Relative weights data for the derivative of $S_P$ with respect to the Fourier component of the Polyakov line
configuration at mode numbers $(210)$.  The underlying lattice gauge theory is the same as in the previous two figures.}
\label{dSk5a}
\end{figure} 

    To see how this goes, we show in Fig.\  \ref{dStheta3k0} our results for
\beq
{1\over L^3} \left( {\pa S_P \over \pa a^R_0} \right)_{a_0=0.03}^{\m/T=i\th} ~~\mbox{vs.}~~ \th
\eeq
together with a best fit of the data to the form
\beq
     f(\th) = c_0 + c_1 \cos(\th)
\eeq
for a lattice gauge theory on a $16^3 \times 4$ lattice volume with $\b=5.2$ (Wilson action) and $ma=0.35$ (unrooted staggered fermions). The fit gives an estimate $h=c_1/6=0.0274(2)$ at $\a=0.03$.  In view of this, we seem to be justified in ignoring terms of order $h^2$ and higher in eqs.\ \rf{deriv1} and \rf{deriv2}.  The data for $h$ is collected at several values of $\a$, and then
extrapolated to $\a=0$, as shown in Fig.\ \ref{h}.
The constant $c_0$ gives an estimate for $\tK(0)$, and this can also be extrapolated to $\a=0$.  For $\vk \ne 0$ we may dispense with the imaginary chemical potential, and simply compute the left hand side of \rf{deriv2} at $\th=0$ at selected values of $\a$.  A typical result is shown in Fig.\ \ref{dSk5a}.
for the mode triplet $(m_1 m_2 m_3) = (210)$.  
From the slope of a best straight-line fit through the data, we determine $\tK(\vk)$ at this particular wavevector.

   For the results shown in the next section, $h$ and $\tK(\vk)$ have been determined by the procedure just described.

\section{Results for the PLA}

   In this initial study we have concentrated on parameters ($\b$, quark mass $ma$, and inverse temperature $N_t$ in lattice units) which bring us close to, but not past, the deconfinement
transition.  In all cases we work on a $16^3 \times N_t$ lattice with staggered, unrooted fermions.

\subsection{Wilson action, $N_t=4$}

\subsubsection{$\b=5.04,~ma=0.2$}

\begin{figure}[htb]
\subfigure[]  
{   
 \label{ks504}
 \includegraphics[scale=0.6]{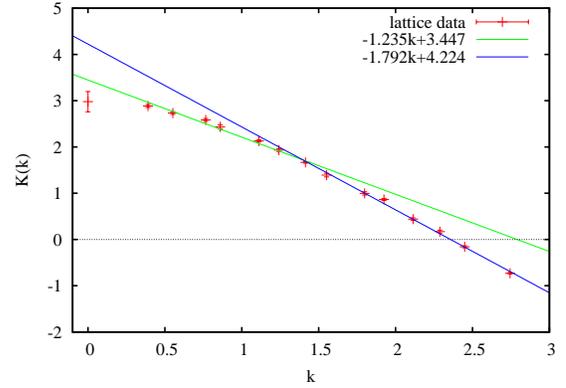}
}
\subfigure[]  
{   
 \label{kk504}
 \includegraphics[scale=0.6]{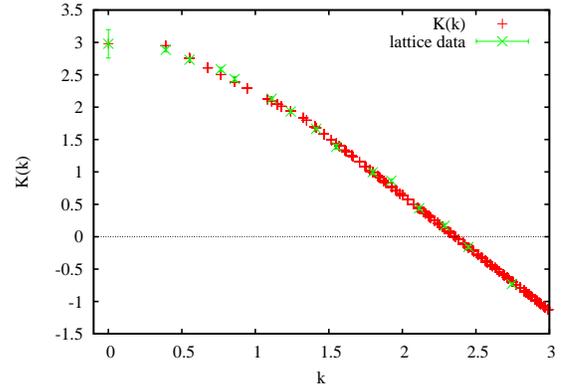}
}
\caption{For the lattice gauge theory at $\b=5.04,~ma=0.2,~N_t=4$: (a) Relative weights results for $\tK(\vk)$ vs $k_L$.  Most of the data points are fit by the two straight lines
shown.  (b) The two straight line fits for $\tK(\vk)$, combined with a long-range cutoff, results in the computed value
$\tK^{fit}$ which also fits the data point at $\vk=0$.}
\label{k504}
\end{figure}

\begin{figure}[htb]
\centerline{\scalebox{0.6}{\includegraphics{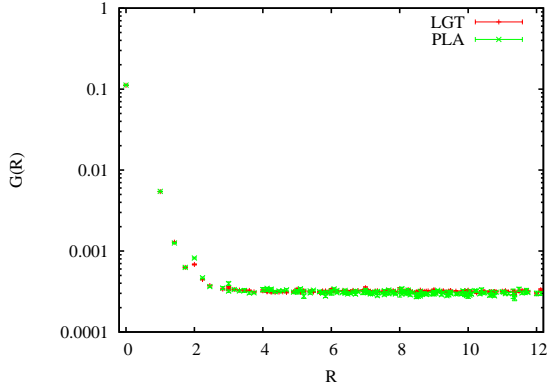}}}
\caption{Comparison of Polyakov line correlators $G(R)$ vs.\ $R$ computed in the lattice gauge theory at 
$\b=5.04,~ma=0.2,~N_t=4$, and in the corresponding PLA derived via relative weights.}
\label{plcorr504}
\end{figure}

   Figure \ref{ks504} is a plot of $K(\vk)$ vs.\ the lattice momentum
\beq
           k_L = 2 \sqrt{ \sum_{i=1}^3 \sin^2(k_i/2)} 
\eeq
We have found in previous work \cite{Greensite:2014isa}, and find here also, that most of the data points can be fit by two straight
lines
\bea
            \tK^{fit}(\vk) = \left\{ \begin{array}{cl}
                   c_1 - c_2 k_L & k_L \le k_0 \cr
                   d_1 - d_2 k_L & k_L \ge k_0 \end{array} \right.
\label{Kfit}
\eea
The exception is one or two points at the lowest momentum, which do not fall on a straight line.  If in fact $\tK^{fit}(\vk)$
would fit $\tK(\vk)$ down to $k_L=0$, it would imply in position space that $K(\vx-\vy) \propto 1/|\vx-\vy|^4$.  As in previous work, we interpret the deviation as implying a cutoff on the long range couplings, and define the position-space kernel with a long distance cutoff $r_{max}$
\beq
    K(\vx-\vy) = \left\{ \begin{array}{cl}
                   {1\over L^3}\sum_\vk \tK^{fit}(k_L) e^{i\vk\cdot (\vx-\vy)} & |\vx-\vy| \le r_{max} \cr \\
                      0 & |\vx-\vy| > r_{max} \end{array} \right. \ .
\eeq
The cutoff $r_{max}$ is chosen so that, upon transforming this kernel back to momentum space, the resulting $\tK(k)$ also fits the low-momentum data at low momenta.  The result of this procedure is shown in Fig.\ \ref{kk504}.  

   The constant $h=0.033$ is determined as explained in the previous section.  The parameter $h$ and the kernel 
$K(\vx-\vy)$ are sufficient to specify the PLA, assuming the validity of the heavy-quark ansatz \rf{eq:SP}, and at
zero chemical potential we may simulate both the PLA and the underlying lattice gauge theory (LGT) to compute and compare  the Polyakov line correlators in each theory.  The result is shown in Fig.\ \ref{plcorr504}

\subsubsection{$\b=5.2,~ma=0.35$}

\begin{figure}[htb]
\subfigure[]  
{   
 \label{ks52}
 \includegraphics[scale=0.6]{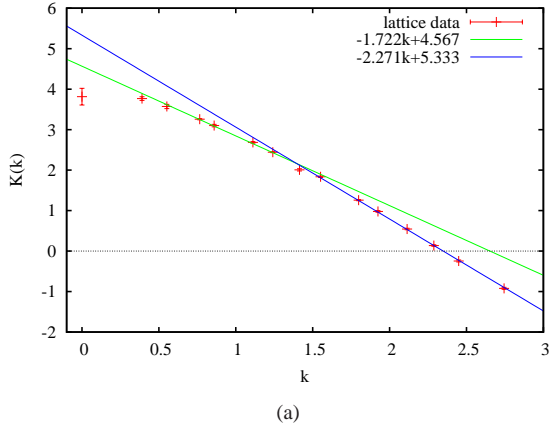}
}
\subfigure[]  
{   
 \label{plcorr52}
 \includegraphics[scale=0.6]{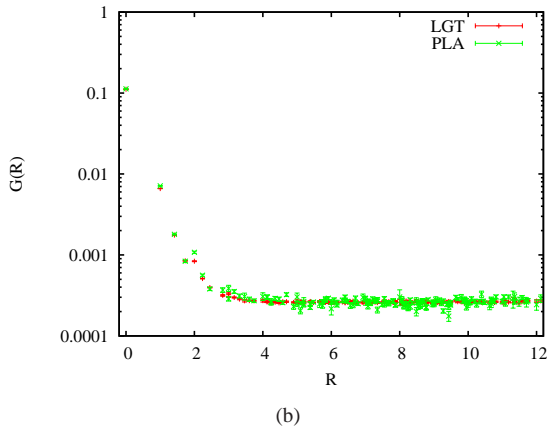}
}
\caption{(a) same as Fig.\ \ref{ks504}; and (b) same as Fig.\ \ref{plcorr504}, for the underlying lattice gauge
theory with $\b=5.2,~ma=0.35,~N_t=4$.}
\label{f52}
\end{figure}

   Fig.\ \ref{ks52} is a plot of $\tK(\vk)$ vs $k_L$, and the analysis proceeds as in the previous section.  The comparison
of Polyakov line correlators in the PLA and LGT is shown in Fig.\ \ref{plcorr52}.
   
\subsubsection{$\b=5.4,~ma=0.6$}

\begin{figure}[htb]
\subfigure[]  
{   
 \label{ks54}
 \includegraphics[scale=0.6]{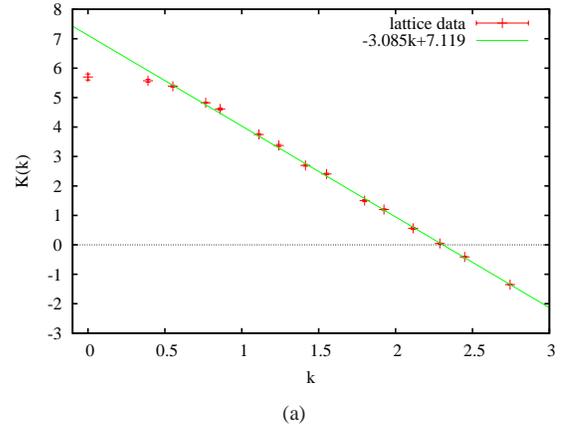}
}
\subfigure[]  
{   
 \label{plcorr54}
 \includegraphics[scale=0.6]{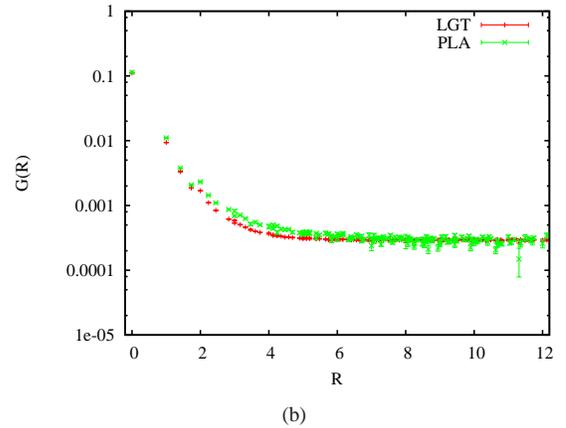}
}
\caption{(a) same as Fig.\ \ref{ks504}; and (b) same as Fig.\ \ref{plcorr504}, for the underlying lattice gauge
theory with $\b=5.4,~ma=0.6,~N_t=4$.}
\label{f54}
\end{figure}

   Plots of of $\tK(\vk)$ vs $k_L$ and the comparison of Polyakov line correlators are shown in Figs.\ \ref{ks54} and
\ref{plcorr54} respectively.

\subsection{L\"uscher-Weisz action, $N_t=6, ~\b=7.0,~ma=0.3$}

\begin{figure}[htb]
\subfigure[]  
{   
 \label{ks70}
 \includegraphics[scale=0.6]{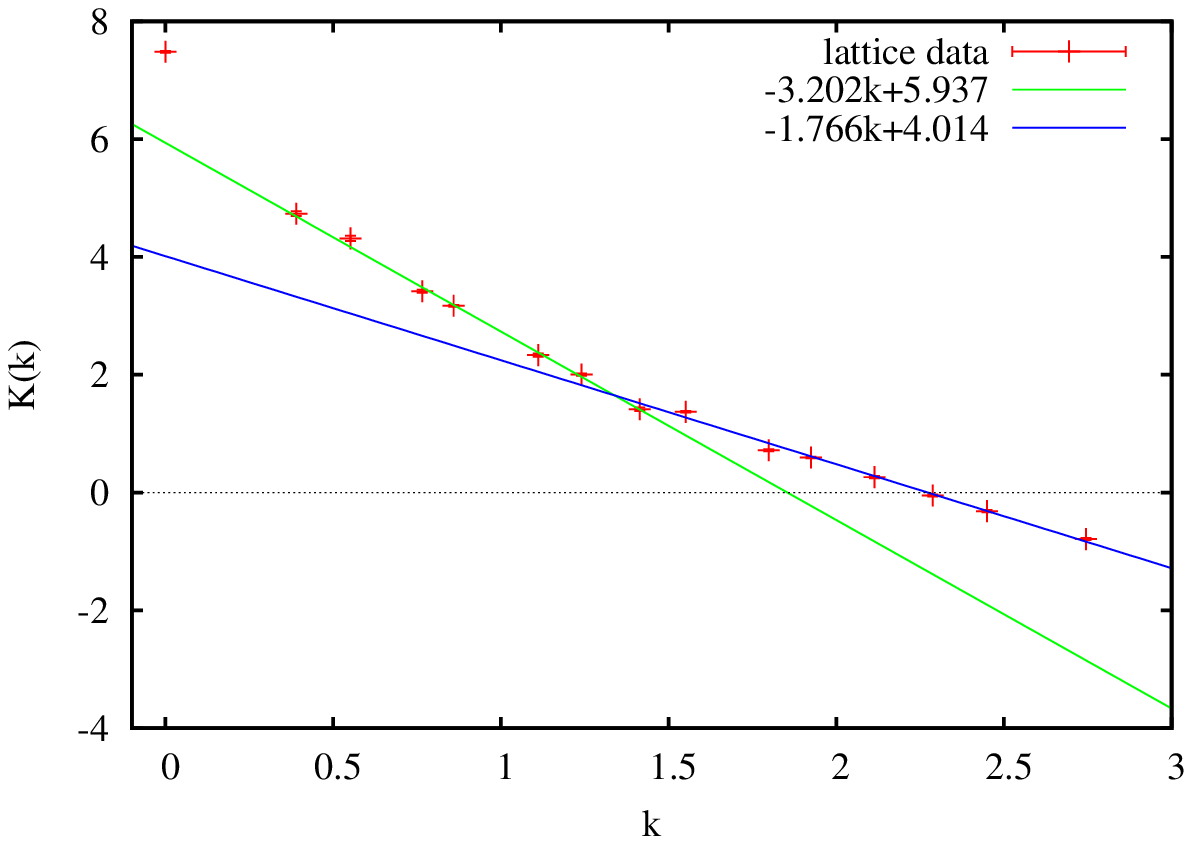}
}
\subfigure[]  
{   
 \label{plcorr70}
 \includegraphics[scale=0.6]{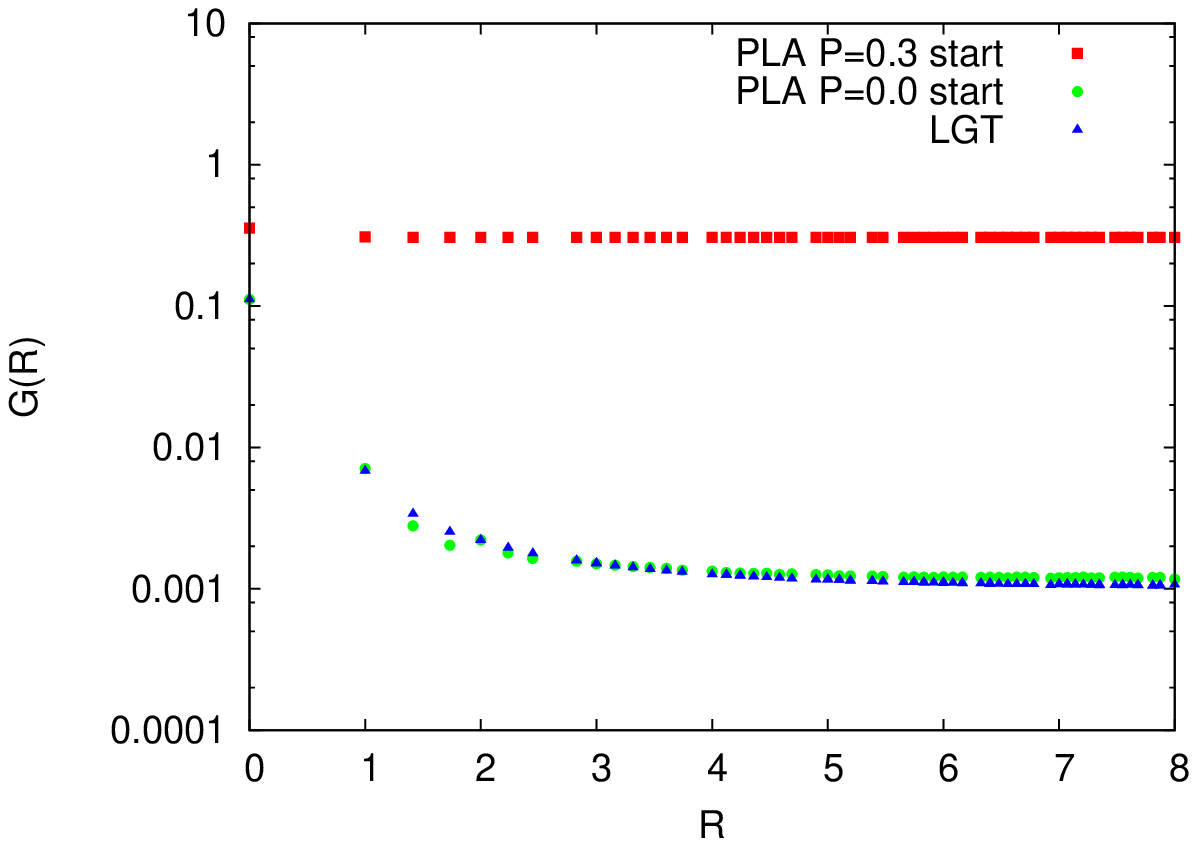}
}
\caption{(a) same as Fig.\ \ref{ks504}; and (b) same as Fig.\ \ref{plcorr504}, for the underlying lattice gauge
theory with the L\"uscher-Weisz action at $\b=7,~ma=0.3,~N_t=6$.  In this case the result for the Polyakov line correlator determined by numerical simulation of the effective action depends on the initialization.  Upper data in (b) is obtained by
initializing at $P_\vx=0.3$, and the lower data is obtained by initialization at $P_\vx=0$.  The lower data points agree quite well with $G(R)$ computed in the underlying lattice gauge theory, which are also shown.}
\label{f70}
\end{figure}

   We have also applied the relative weights method to the L\"uscher-Weisz action, with the parameters listed above.
Again most of the $\tK(\vk)$ data points can be fit by two straight lines.  However, there is a significant difference as compared to the previous cases at
$\vk=0$, where $\tK(0)$ lies above, rather than below the straight line, as seen in Fig.\ \ref{ks70}.  Neglecting couplings between lattice sites beyond some separation $r_{max}$ will inevitably result in disagreement with the $\tK(0)$ data point.

\begin{table*}[t!]
\begin{center}
\begin{tabular}{|c|c|c|c|c|c|c|c|c|c|c|} \hline
         action &  $ N_t $ & $\b$ & $ma$ & $c_1$ &  $c_2$ & $d_1$ & $d_2$ & $r_{max}$ & h  \\
\hline
        Wilson &         4 &  5.04 & 0.2  & 3.45 & 1.24  & 4.22 & 1.79 & 4.2  & 0.0334 \\
        Wilson &         4 &  5.2  & 0.35 & 4.57 & 1.72  & 5.33 & 2.27 & 2.3  & 0.0264  \\ 
        Wilson &         4 &  5.4  & 0.6   & 7.12 &  3.09 &   $-$ &  $-$  &  3.4 & 0.0168 \\ 
L\"uscher-Weisz &  6 &  7.0 & 0.3    & 5.94 & 3.20  & 4.01 & 1.77 & $\infty$   & 0.0117 \\          
\hline
\end{tabular}
\caption{Parameters defining the effective Polyakov line actions $S_P$, for the corresponding 
 SU(3) lattice gauge theories with dynamical staggered fermions on a $16^3 \times N_t$ lattice volume.
 The lattice gauge theory is specified in the the first four entries on each row, and the effective action used to
 compute Polyakov line correlators is
 described by the remaining parameters.  In the L\"uscher-Weisz case, with $r_{max}=\infty$, it is also necessary
 to specify $\tK(0)= 7.46$ in defining $S_P$, as discussed in the text.} 
\label{tab1}
\end{center}
\end{table*}

   In this case the strategy is to Fourier transform the two-line fit \rf{Kfit} to position space, with the modification
that we identify $K^{fit}(0)$ with $\tK(0)$, and dispense with a finite-distance cutoff at $r_{max}$.  The resulting kernel
$K(\vx-\vy)$ in the PLA couples each lattice site to every other lattice site.  The result appears to be multiple metastable phases, which depend, in numerical simulations, on the initial configuration.

   In Fig.\ \ref{plcorr70} we display our results for the Polyakov line correlator $G(R)$ obtained from numerical simulations of 
\begin{itemize}
\item the Polyakov line action with a starting configuration initialized to $P_\vx = 0.3$;
\item the Polyakov line action with a starting configuration initialized to $P_\vx = 0.0$;
\item the underlying lattice gauge theory.
\end{itemize}
These results indicate that there are at least two phases in the PLA, confined and deconfined, which are stable over thousands of Monte Carlo sweeps.  The Polyakov line correlator of the PLA in the confined phase is consistent with the correlator in the underlying lattice gauge theory, while the correlator in the deconfined phase is not.  It seems that for the
purpose of numerical simulations the effection action alone may be insufficient, and it may be necessary in some cases to supplement the PLA with a prescription for initialization of the SU(3) spin system.

   The existence of multiple stable or metastable phases in the PLA is very clearly associated with the long-range couplings in the effective action.  We have checked that if one simply truncates the range of bilinear couplings then the multiple phases disappear, and the result for the Polyakov line correlator is independent of the initialization.  Of course, that arbitrary truncation also destroys the agreement of the correlators obtained in the PLA and the underlying lattice gauge theory.
   
\bigskip

   Parameters which describe the effective actions in each of the cases considered above are displayed in Table \ref{tab1}.

\bigskip

\section{Mean field solutions at finite density}

   We review here the mean field approach to solving the PLA at finite density, as explained in refs.\ 
\cite{Greensite:2012xv} and \cite{Greensite:2014cxa}.  The partition function corresponding to the action \rf{eq:SP} is
\bea
   Z &=& \int \prod_\vx dU_\vx \Dt_\vx(\m,\tr U,\tr U^\dg) e^{S_0}
\non \\
   S_0 &=& \sum {1\over 9} K(\vx-\vy) \tr U_\vx \tr U_\vy
\eea
with
\begin{widetext}
\bea
   \Dt_\vx(\m,\tr U,\tr U^\dg)) &=&(1+he^{\mu/T}\tr U_\vx+h^2e^{2\mu/T}\tr
	U^\dg_\vx+h^3e^{3\mu/T})(1+he^{-\mu/T}\tr U^\dg_\vx+h^2e^{-2\mu/T}\tr U_\vx+h^3e^{-3\mu/T}) 
\non \\
&=& a_1 + a_2 \tr U_\vx + a_3 \tr U^\dg_\vx + a_4 (\tr U_\vx)^2 + a_5 (\tr U^\dg_\vx)^2 + a_6 \tr U_\vx \tr U^\dg_\vx
\eea
and
\bea
a_1 &=& 1 + h^3 (e^{3\mu/T}+e^{-3\mu/T}) + h^6
\non \\
a_2 &=&  (1 + h^2)^2 h e^\m + (1 + h^2) h^2 e6{-2\m} - 2 h^3 e^\m~~~,~~~ a_3 =  (h+h^5)e^{-\m/T} + (h^2+h^4)e^{2\m/T} 
\non \\
a_4 &=& h^3 e^{-\m/T} ~~~,~~~ a_5 = h^3 e^{\m/T} ~~~,~~~ a_6 = h^2 + h^3 \ .
\eea
\end{widetext}
We then write
\bea
           S^0_P &=&  {1\over 9} \sum_{(\vx \vy)} \tr[U_\vx] \tr[U_\vy^\dg] K(\vx-\vy) + 
                               a_0 \sum_{\vx} \tr[U_\vx] \tr[U_\vx^\dg] \ ,
\eea
where we introduce the notation for the double sum excluding $\vx=\vy$
\beq
    \sum_{(\vx \vy)} \equiv \sum_{\vx} \sum_{\vy} (1-\d_{\vx,\vy})  \ .
\eeq
Next introduce parameters $u,v$
\beq
   \tr U_\vx = (\tr U_\vx - u) + u  ~~~,~~~ \tr U^\dg_\vx = (\tr U^\dg_\vx - v) + v 
\eeq
so that
\bea
S_0 &=& J_0 \sum_\vx (v \tr U_\vx + u \tr U_\vx^\dg) - uvJ_0V 
\non \\
& &+ a_0 \sum_{\vx} \tr[U_\vx] \tr[U_\vx^\dg]+E_0 \ ,
\eea
where $V=L^3$ is the lattice volume, and we have defined
\bea
     E_0 &=& \sum_{(\vx \vy)} (\tr U_x-u)(\tr U_\vy^\dg - v) \on K(\vx-\vy) \ ,
\non \\
     J_0 &=&  {1\over 9} \sum_{\vx \ne 0} K(\vx)  ~~~,~~~ a_0 = {1\over 9} K(0) \ .
\eea
Parameters $u$ and $v$ are to be chosen such that $E_0$ can be treated as a perturbation, to be ignored as a first approximation. In particular, $\langle E_0 \rangle = 0$ when
\bea
u = \langle \tr U_x \rangle ~~~,~~~ v = \langle \tr U^\dg_x \rangle \ .
\label{consistency}
\eea
These conditions turn out to be equivalent to the stationarity of the mean field free energy.  The leading mean field result is obtained by dropping $E_0$, in which case the integrand of the partition function factorizes
\begin{widetext}
\bea
Z_{mf} &=& e^{-uvJ_0 V} \prod_x \int dU_\vx \Dt_\vx(\m,\tr U, \tr U^\dg) 
 \exp[a_0 \tr U_\vx \tr U^\dg_\vx]  e^{A \tr U_\vx + B \tr U^\dg_\vx}
\non \\
&=& e^{-uvJ_0 V} \bigg\{ \Dt\left(\m,{\pa \over \pa A},{\pa \over \pa B} \right) \exp\left[a_0 {\pa^2 \over \pa A \pa B} \right]
           \int dU e^{A \tr U + B \tr U^\dg} \bigg\}^V
\eea
\end{widetext}
where $A=J_0 v,~B=J_0 u$.  The SU(3) group integral is known (see, e.g., \cite{Greensite:2012xv}),
\bea
 \int dU e^{A \tr U + B \tr U^\dg} &=&  \sum_{s=-\infty}^{\infty}   \det\Bigl[D^{-s}_{ij} I_0[2\sqrt{A B}] \Big] 
\eea
where $D^{-s}_{ij}$ is the $i,j$-th component of a matrix of differential operators
\bea
D^s_{ij} &=& \left\{ \begin{array}{cl}
                         D_{i,j+s} & s \ge 0 \cr
                         D_{i+|s|,j} & s < 0 \end{array} \right. \ ,
\non \\
D_{ij} &=& \left\{ \begin{array}{cl}
                         \left({\pa \over \pa B} \right)^{i-j} & i \ge j \cr 
                        \left({\pa \over \pa A} \right)^{j-i} & i < j \end{array} \right. \ ,
\eea
Putting everything together, with $Z_{mf} = \exp[ -f_{mf}V/T]$, the mean-field free energy/volume is
\beq
{f_{mf} \over T} = J_0 u v - \log G(A,B)
\eeq
where
\bea
G(A,B) =  \Dt\left(\m,{\pa \over \pa A},{\pa \over \pa B} \right) 
 \sum_{s=-\infty}^{\infty}   \det\Bigl[D^{-s}_{ij} I_0[2\sqrt{A B}] \Big] 
\label{G}
\eea

With these definitions, the mean field values
\beq
\langle \tr U \rangle = u ~~~,~~~ \langle \tr U^\dg \rangle = v
\eeq
are obtained from the solution of the simultaneous equations
\bea
u - {1\over G}{\pa G \over \pa A}=0  ~~~~~\text{and}~~~~~ v - {1\over G}{\pa G \over \pa B}=0 \ ,
\label{hq-conditions}
\eea
with the number density given by
\bea
           n &=& {1\over G} {\pa G \over \pa \m}
\eea
In practice a computation of the mean field
estimate $f_{mf}$ of the free energy requires a truncation of the sum over $s$ in \rf{G}, an expansion in $a_0$ to finite order, and a check that the results are not sensitive to increasing the cutoff.  We have found that restricting the sum over $s$ to the range $-3 \le s \le 3$, and the expansion to $a_0$ to second order, is sufficient.

\begin{figure}[htb]
\subfigure[]  
{   
 \label{u504}
 \includegraphics[scale=0.6]{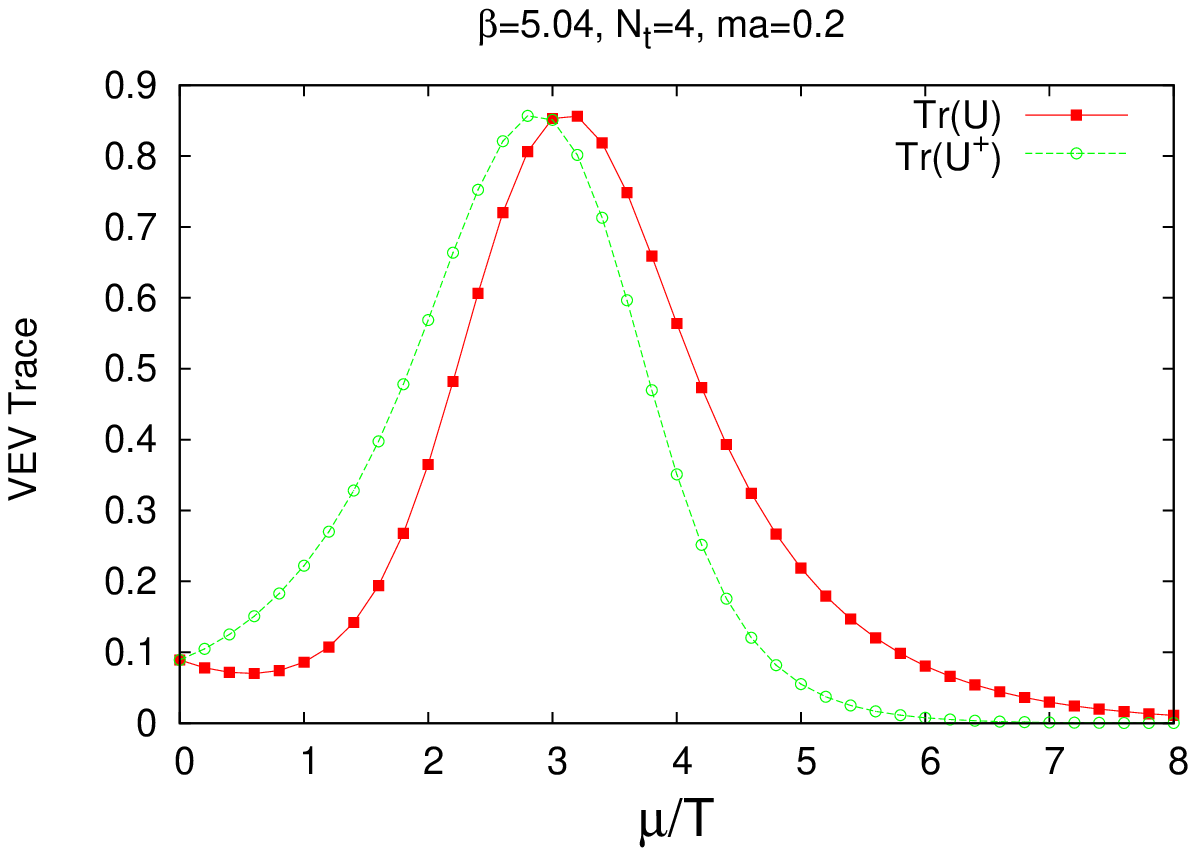}
}
\subfigure[]  
{   
 \label{n504}
 \includegraphics[scale=0.6]{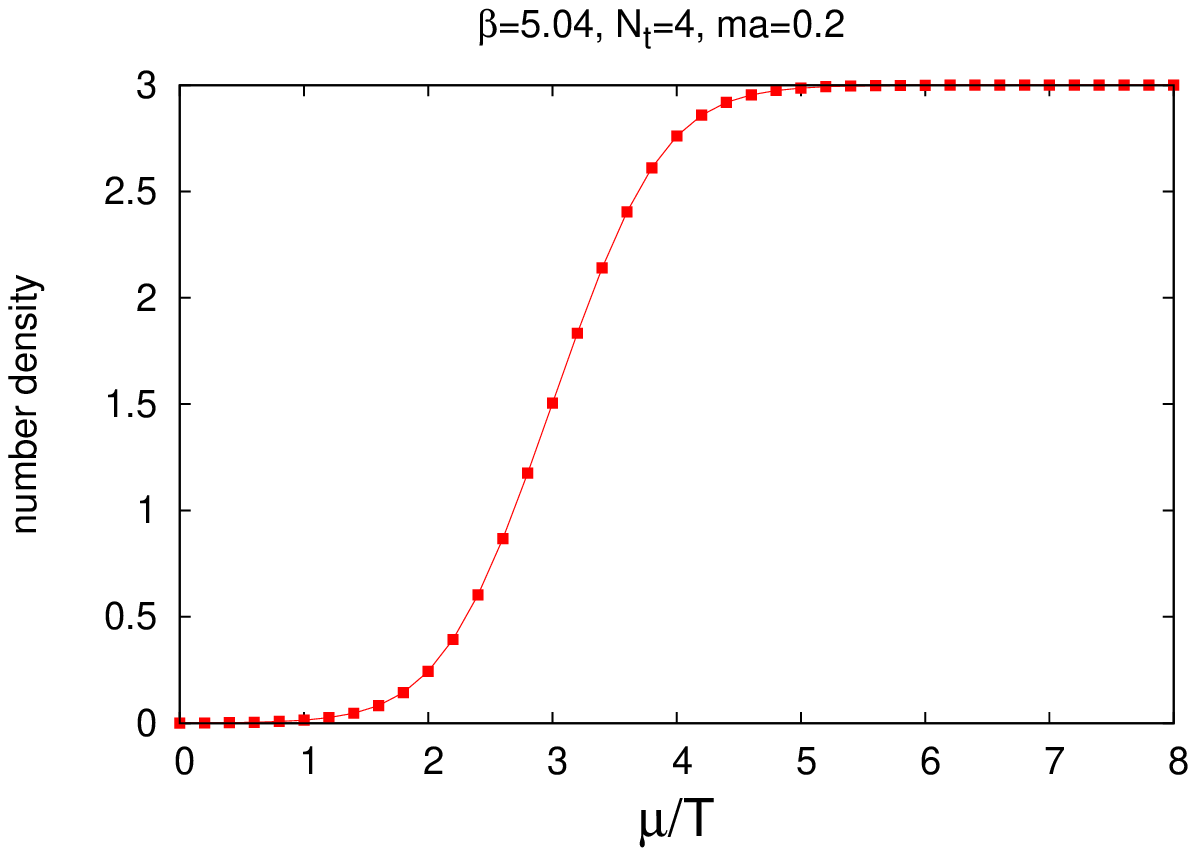}
}
\caption{Mean field solution of the PLA corresponding to a Wilson action lattice gauge
theory at $\b=5.04,~ma=0.2,~N_t=4$ at finite density $\m$.  (a) Expectation values of $\tr U,~\tr U^\dg$ vs.\ $\m$.
(b) Quark number density $n$ vs.\ $\m$.}
\label{mf504}
\end{figure}

\begin{figure}[htb]
\subfigure[]  
{   
 \label{u540}
 \includegraphics[scale=0.6]{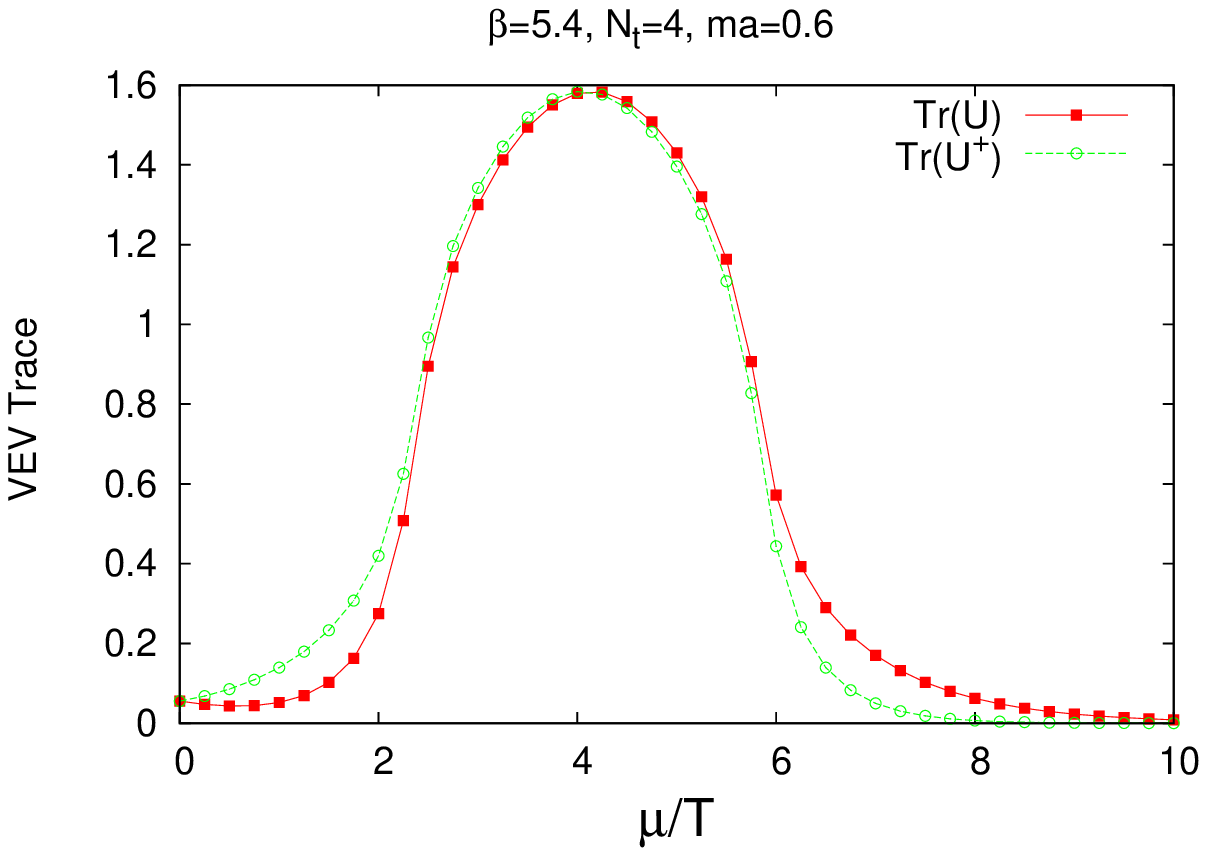}
}
\subfigure[]  
{   
 \label{n540}
 \includegraphics[scale=0.6]{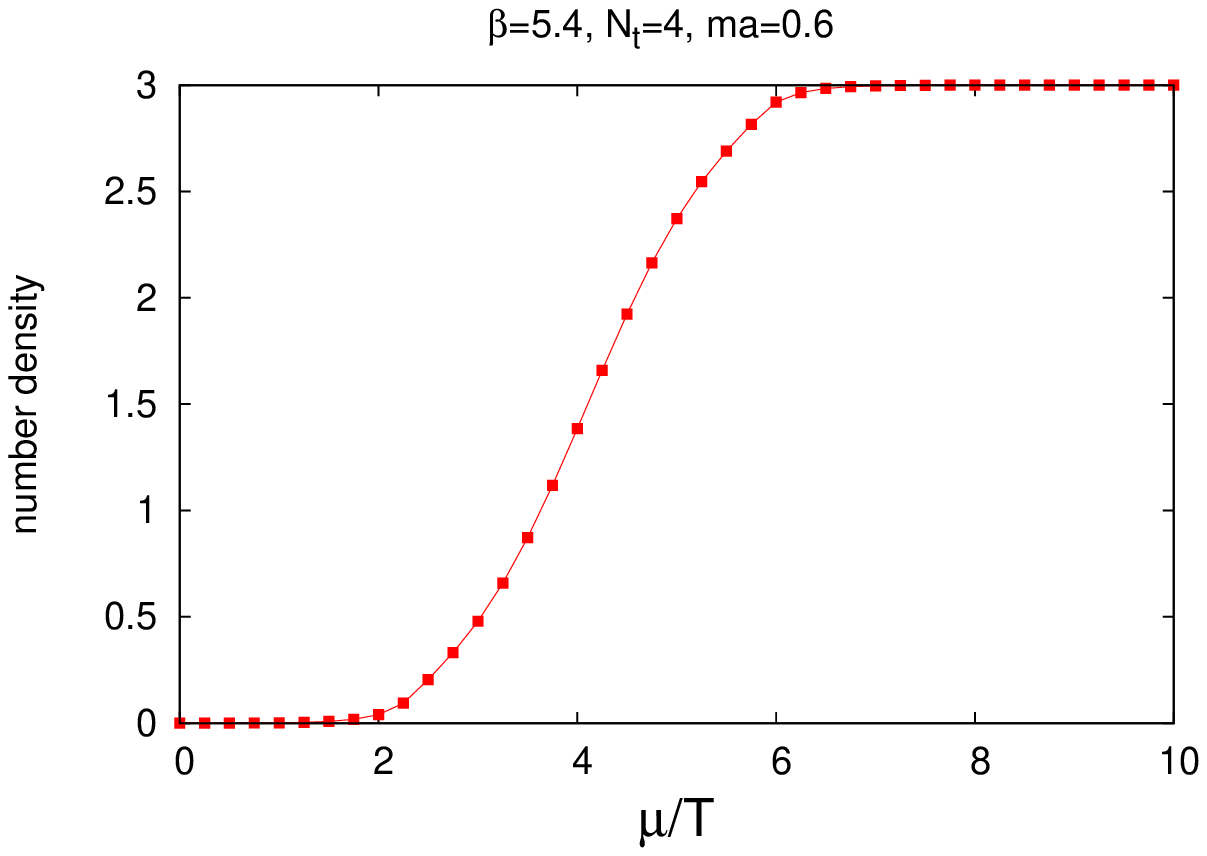}
}
\caption{Mean field solution of the PLA corresponding to a Wilson action lattice gauge
theory at $\b=5.4,~ma=0.6,~N_t=4$ at finite density $\m$.  (a) Expectation values of $\tr U,~\tr U^\dg$ vs.\ $\m$.
(b) Quark number density $n$ vs.\ $\m$.}
\label{mf540}
\end{figure}

   The results for the examples we have considered in the last section, with the Wilson action and $N_t=4$, are qualitatively very much like the mean field results heavy quark cases, which were reported in \cite{Greensite:2014cxa}.  The mean field solutions for $\langle \tr U \rangle, \langle \tr U^\dg \rangle$ and number density $n$ for the cases $\b=5.04,~ma=0.2$
and $\b=5.4,~ma=0.6$ are shown in Figs.\ \ref{mf504} and \ref{mf540}.
   
\begin{figure*}[htb]
\subfigure[]  
{   
 \label{u70}
 \includegraphics[scale=0.6]{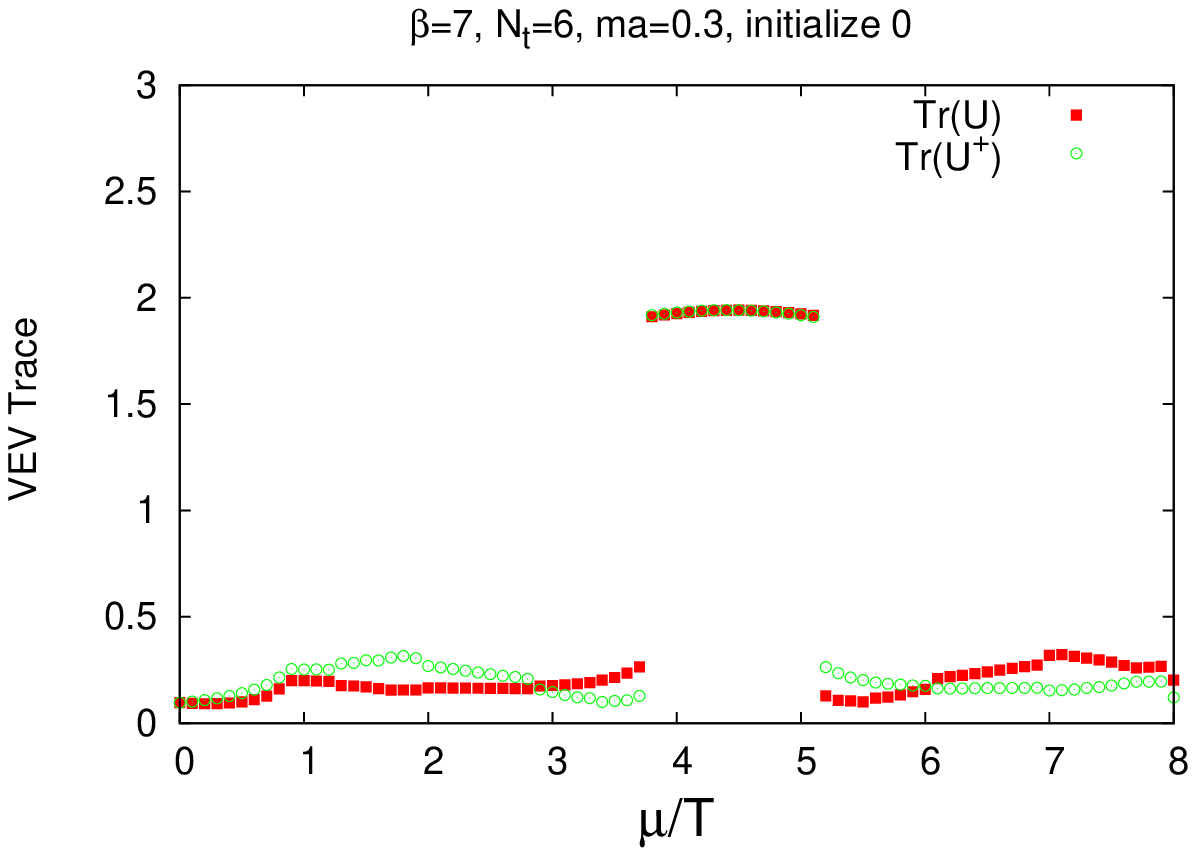}
}
\subfigure[]  
{   
 \label{n70}
 \includegraphics[scale=0.6]{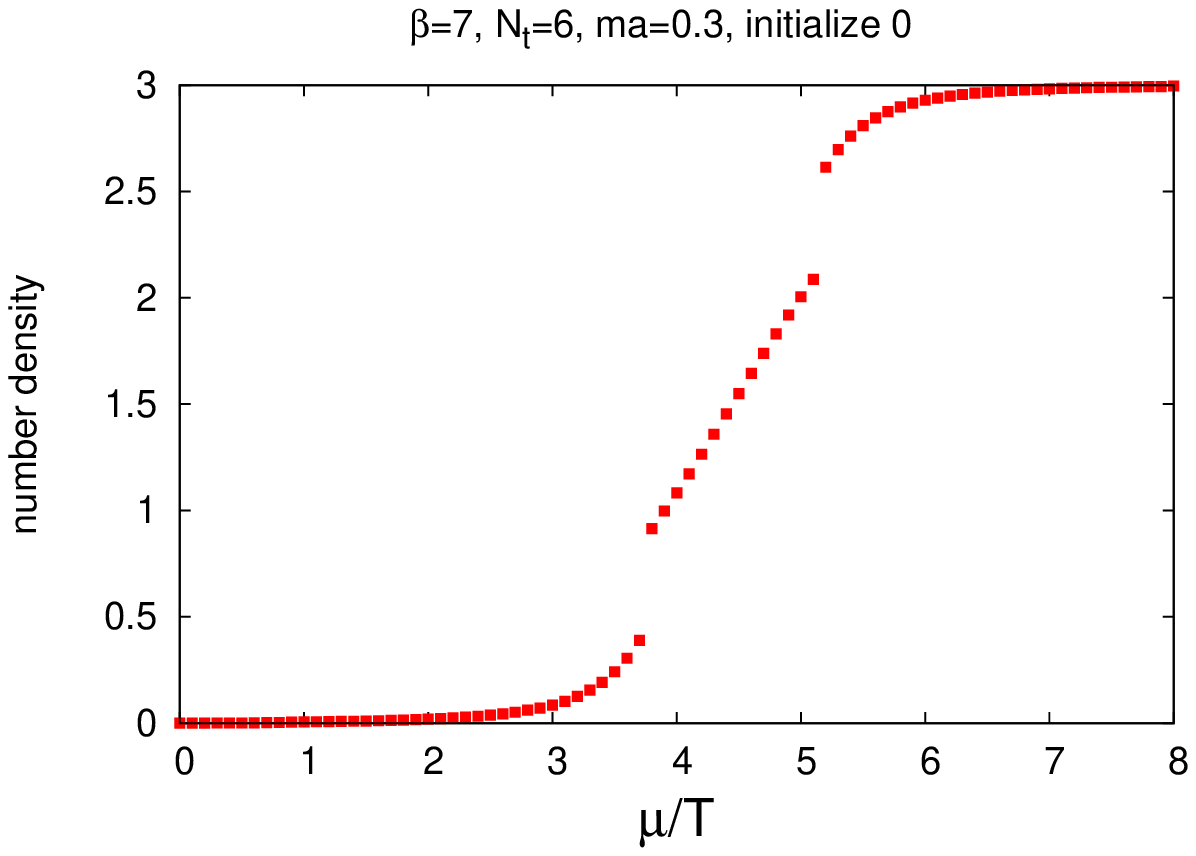}
}
\caption{A mean field solution of the PLA corresponding to a L\"uscher-Weisz action lattice gauge
theory at $\b=7.0,~ma=0.3,~N_t=6$ at finite density $\m$.  In this case the routines look for a solution of the mean field equations \rf{hq-conditions} closest to $u=v=0$.  (a) Expectation values of $\tr U,~\tr U^\dg$ vs.\ $\m$.
(b) Quark number density $n$ vs.\ $\m$.}
\label{mf70a}
\end{figure*}

\begin{figure}[htb]
\subfigure[]  
{   
 \label{u70b}
 \includegraphics[scale=0.6]{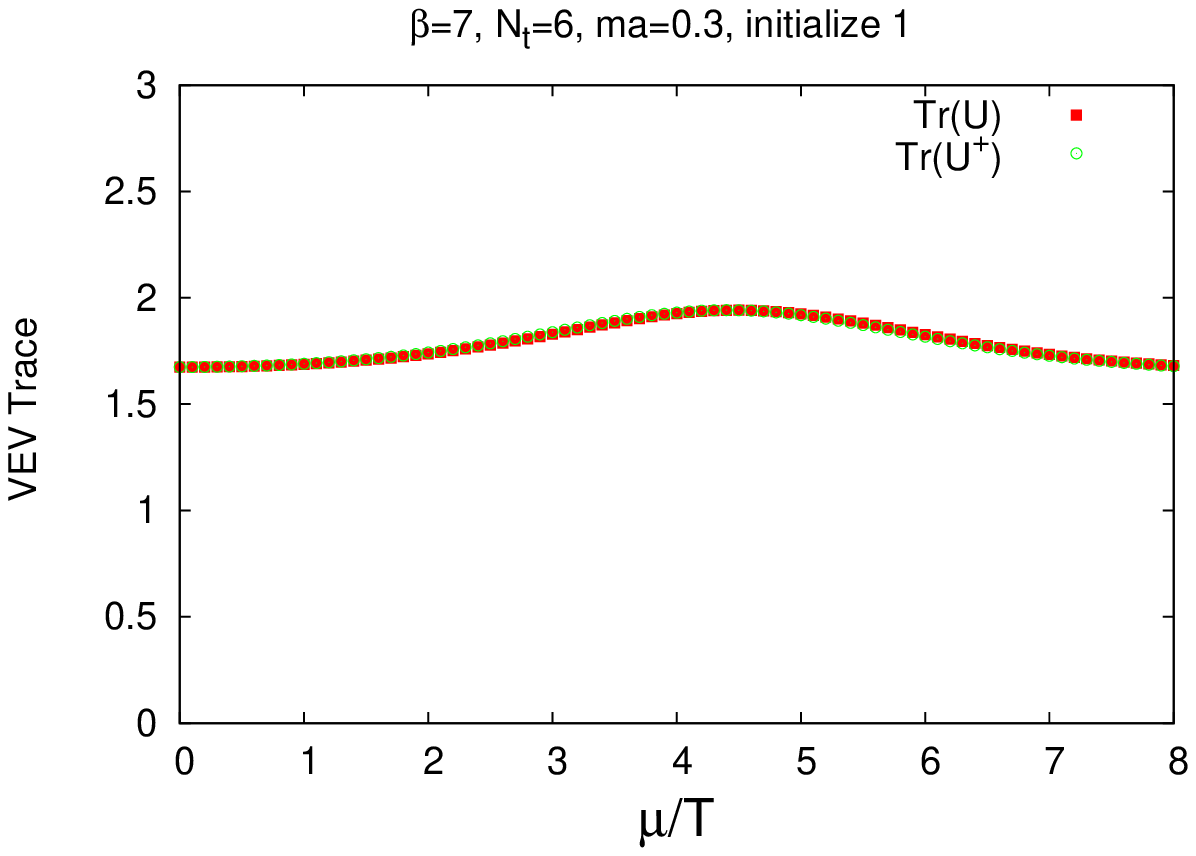}
}
\subfigure[]  
{   
 \label{n70b}
 \includegraphics[scale=0.6]{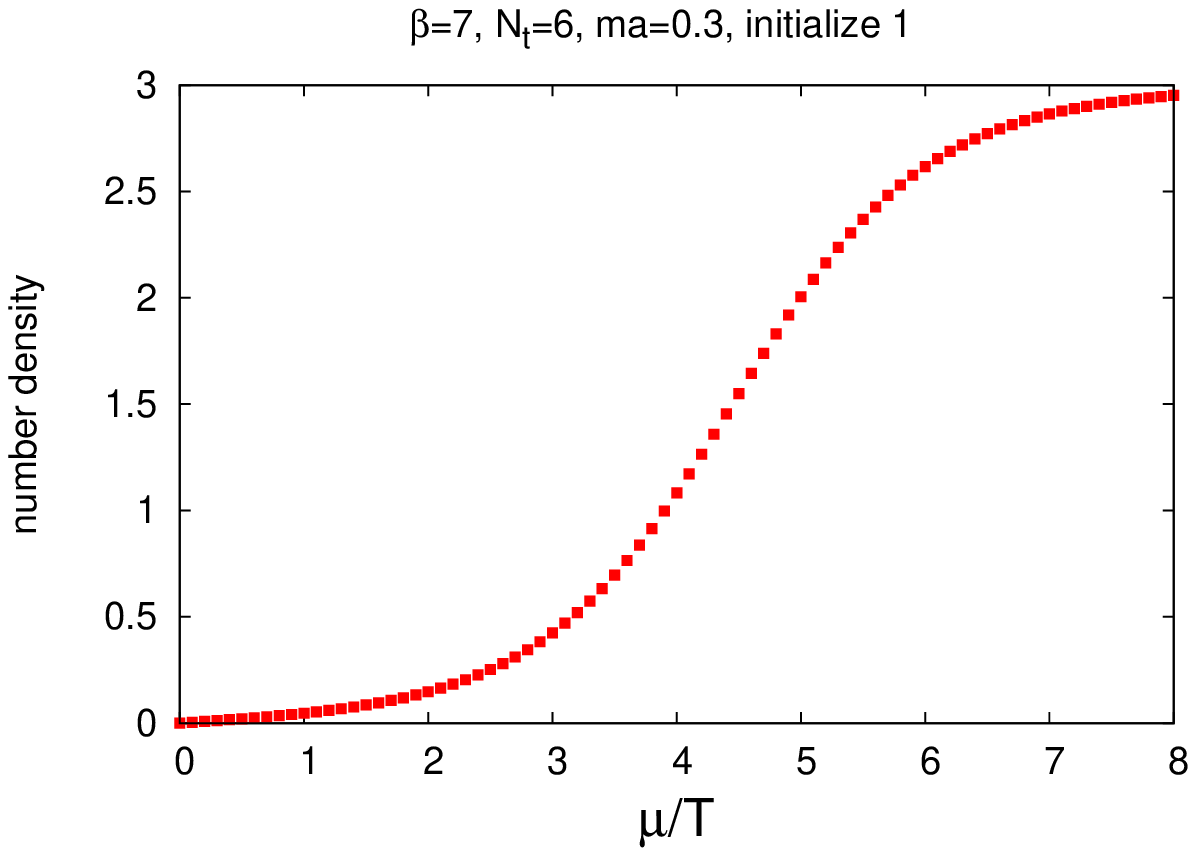}
}
\caption{A mean field solution of the PLA corresponding to a L\"uscher-Weisz action lattice gauge
theory at $\b=7.0,~ma=0.3,~N_t=6$ at finite density $\m$.  In this case the routines look for a solution of the mean field equations \rf{hq-conditions} closest to $u=v=1$.  (a) Expectation values of $\tr U,~\tr U^\dg$ vs.\ $\m$.
(b) Quark number density $n$ vs.\ $\m$.}
\label{mf70b}
\end{figure}

\begin{figure}[t!]
\centerline{\scalebox{0.6}{\includegraphics{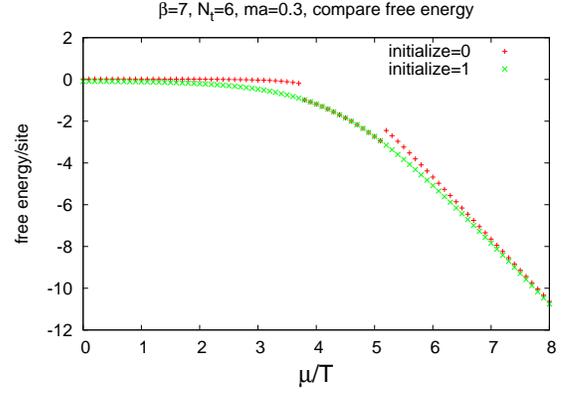}}}
\caption{The mean field free energy corresponding to solutions shown in the previous two figures.  Where the solutions differ, the solutions with larger $\tr U,~\tr U^\dg$ have the lower free energy.}
\label{free}
\end{figure}

   The L\"uscher-Weisz action at $N_t=6,~ \b=7.0,~ ma=0.3$ is more interesting.  There are multiple solutions of the mean-field equations \rf{hq-conditions}, and the solution which is found by a search routine depends on the starting values for
$u$ and $v$.   Initialization at $u=v$ near zero gives the results 
shown in Fig.\  \ref{mf70a}.   Here there seem to be two clear phase transitions at finite density.
If, however, the search routines begin at $u=v=1$, then solutions correspond to the deconfined phase at $\m=0$, and
there is no transition found at any value of $\m$, as seen in Fig.\ \ref{mf70b}.  Ordinarily the stable phase corresponds to the phase with lowest free energy, and by this criterion (see Fig.\ \ref{free}) the solutions shown in Fig.\ \ref{mf70b} are 
selected. However, we have found that at $\m=0$ this is {\it not} the phase which corresponds to the phase of the underlying lattice gauge theory.  This of course raises the question of which metastable state corresponds to the state of the underlying gauge theory at finite density.   

\begin{table*}[htb]
\begin{center}
\begin{tabular}{|c|c|c|c|c|c|} \hline
         action &  $ N_t $ & $\b$ & $ma$ &${1\over 3}\langle \tr U \rangle$& ${1\over 3}\langle \tr U \rangle_{mf}$ \\
\hline
        Wilson &          4 &  5.04 & 0.2  & 0.01778(3) & 0.01765   \\
        Wilson &          4 &  5.2  & 0.35 & 0.01612(4) & 0.01603    \\ 
        Wilson &          4 &  5.4  & 0.6   & 0.01709(5) &  0.01842  \\ 
L\"uscher-Weisz I&  6 &  7.0 & 0.3    & 0.03580(4) & 0.03212    \\   
L\"uscher-Weisz II& 6 &  7.0 & 0.3    & 0.554(1) & 0.5580    \\         
\hline
\end{tabular}
\caption{Polyakov line expectation values from numerical simulations of lattice gauge theory (column 5) , compared to mean field estimates (column 6).  For the L\"uscher-Weisz action there are multiple solutions of the mean field equations. The solution in L\"uscher-Weisz I is the one found by a search routine initialized at $u=v=0$, while the solution in L\"uscher-Weisz II corresponds to initialization at $u=v=1$. For L\"uscher-Weisz II, the value in column 5 was obtained from numerical simulation of the PLA, rather than the lattice gauge theory, with Polyakov lines initialized to 0.3.}
\label{tab2}
\end{center}
\end{table*}

\subsection{Validity of Mean Field at $\m=0$}

   The mean field method is an approximation technique whose validity depends on each spin being coupled to
many other spins, and for this reason the mean field approach is often thought of as a $1/d$ expansion, with $d$ the number of dimensions.  At least, this is the case for theories with mainly nearest-neighbor couplings.  However, it is
clear that the effective Polyakov line actions corresponding to lattice gauge theories couple each SU(3) spin to a very large number of other spins, and in one case we have looked at (with the L\"uscher-Weisz action) each spin is coupled to all other spins on the lattice.  This means that even in $D=3$ dimensions the mean field method may be quantitatively more accurate then one might naively expect.
One place we can check this is at $\m=0$, where $\langle \tr U \rangle$ can be computed in the underlying lattice gauge theory, and also from the mean field solution of the effective Polyakov line actions.  It turns out that these values are in very accurate agreement, as can be seen in Table \ref{tab2}.   

    In an earlier work \cite{Greensite:2014cxa} we compared the mean-field solution of effective Polyakov line actions corresponding to gauge-Higgs theories, at $\mu \ne 0$, to the corresponding solution of the effective theories by the Langevin equation.   In that work it was found that even at $\m \ne 0$ the mean field results were virtually identical to 
the Langevin results, in every region where the latter could be trusted.  This is in accord with the accuracy we have found for mean field at $\m=0$ with dynamical fermions.
    
\section{Conclusions}

    We have derived effective Polyakov line actions via the relative weights method for several cases of SU(3)
lattice gauge theory with dynamical staggered fermions, and solved these theories at non-zero chemical potential by a 
mean field approach.  At $\m=0$ we find good agreement for the Polyakov line correlators computed in the effective theories and the underlying lattice gauge theories.  We have also found, at $\m=0$, that Polyakov line expectation values computed via mean field theory are in remarkably close agreement with the values obtained by numerical simulation, and this is probably due to the fact that each SU(3) spin is coupled to very many other spins in the effective theory, which favors the mean field approach. 

   However, this non-local feature of the effective action also leads, in the most non-local case we have looked at (each spin coupled to all spins) to an unpleasant feature, namely, the existence of more than one metastable phase.  These phases depend on the initialization chosen, and they persist throughout the numerical simulation, involving thousands of Monte Carlo sweeps.  Since this is a phenomenon seen at $\m=0$, it is not specifically tied to the sign problem, but rather to the non-locality of the effective action in certain cases.  At $\mu \ne 0$ one must either find some criterion for choosing the phase which corresponds to lattice gauge theory, or else restrict the
analysis of the Polyakov line action to cases where the couplings are comparatively short range. It should be emphasized that even if there are significant terms in the action which are ignored in the simple ansatz \rf{eq:SP}, and even if such terms were taken into account, there may still be multiple metastable phases if the bilinear couplings are long (or infinite) range.   Whether, in such cases, some other simulation method (e.g.\ Langevin) could avoid the dependence of phase on initialization remains to be seen.
It is also important to discover whether very long or infinite range couplings in the effective action are generic at small quark mass and small lattice spacings, or whether such cases are special and can be bypassed.

\acknowledgments{JG's research is supported by the U.S.\ Department of Energy under Grant No.\ DE-SC0013682.  RH's research is supported by the Erwin Schr\"odinger Fellowship program of the Austrian Science Fund FWF (``Fonds zur F\"orderung der wissenschaftlichen Forschung'') under Contract No. J3425-N27.}     
   
\bibliography{pline}

\begin{thebibliography}{10}

\bibitem{Langelage:2014vpa}
J.~Langelage, M.~Neuman, and O.~Philipsen,
\newblock JHEP {\bf 09}, 131 (2014), arXiv:1403.4162.

\bibitem{Sexty:2013ica}
D.~Sexty,
\newblock Phys. Lett. {\bf B729}, 108 (2014), arXiv:1307.7748.

\bibitem{Scorzato:2015qts}
L.~Scorzato,
\newblock {The Lefschetz thimble and the sign problem},
\newblock in {\em {Proceedings of Lattice 2015}}, 2015, arXiv:1512.08039.

\bibitem{Greensite:2014isa}
J.~Greensite and K.~Langfeld,
\newblock Phys. Rev. {\bf D90}, 014507 (2014), arXiv:1403.5844.

\bibitem{Greensite:2013yd}
J.~Greensite and K.~Langfeld,
\newblock Phys.Rev. {\bf D87}, 094501 (2013), arXiv:1301.4977.

\bibitem{Bergner:2015rza}
G.~Bergner, J.~Langelage, and O.~Philipsen,
\newblock JHEP {\bf 11}, 010 (2015), arXiv:1505.01021.

\bibitem{Greensite:2013bya}
J.~Greensite and K.~Langfeld,
\newblock Phys.Rev. {\bf D88}, 074503 (2013), arXiv:1305.0048.

\bibitem{Fromm:2011qi}
M.~Fromm, J.~Langelage, S.~Lottini, and O.~Philipsen,
\newblock JHEP {\bf 1201}, 042 (2012), arXiv:1111.4953.

\bibitem{Greensite:2012xv}
J.~Greensite and K.~Splittorff,
\newblock Phys.Rev. {\bf D86}, 074501 (2012), arXiv:1206.1159.

\bibitem{Greensite:2014cxa}
J.~Greensite,
\newblock Phys. Rev. {\bf D90}, 114507 (2014), arXiv:1406.4558.

\end{thebibliography}
 
\end{document}